\documentclass{jfm}

\usepackage{graphicx}
\usepackage{newtxtext}
\usepackage{newtxmath}
\usepackage{natbib}
\usepackage{hyperref}
\hypersetup{
    colorlinks = true,
    urlcolor   = blue,
    citecolor  = black,
}

\newcommand{\RomanNumeralCaps}[1]
\linenumbers
\usepackage{comment}
\usepackage{color}
\usepackage{subcaption}
\usepackage{bm}% bold math

\usepackage[OMLmathsfit,sfdefault=cmbr]{isomath}

\newcommand{\Wi}{\textit{Wi}}
\newcommand{\Sc}{\textit{Sc}}

\shorttitle{EIT: ROM based on manifold dynamics}
\shortauthor{M. Kumar, C. R. Constante-Amores and M. D. Graham}

\title{Elastoinertial turbulence: Data-driven reduced-order model based on manifold dynamics}

\author{Manish Kumar\aff{1}, C. Ricardo Constante-Amores\aff{1,2}, \and Michael D. Graham\aff{1}  \corresp{\email{mdgraham@wisc.edu}}}

\affiliation{\aff{1}Department of Chemical and Biological Engineering, University of Wisconsin-Madison, 1415 Engineering Dr, Madison, WI 53706, USA
\aff{2}Department of Mechanical Science and Engineering, University of Illinois, Urbana Champaign, IL 61801, USA
\\[\affilskip]
}

\begin{document}

\maketitle
\begin{abstract}
Elastoinertial turbulence (EIT) is a chaotic state that emerges in the flows of dilute polymer solutions. Direct numerical simulation (DNS) of EIT is highly computationally expensive due to the need to resolve the multi-scale nature of the system. While DNS of 2D EIT  typically requires $O(10^6)$ degrees of freedom, we demonstrate here that a data-driven modeling framework allows for the construction of an accurate model with 50 degrees of freedom. We achieve a low-dimensional representation of the full state by first applying a viscoelastic variant of proper orthogonal decomposition to DNS results, and then using an autoencoder. The dynamics of this low-dimensional representation are learned using the neural ODE method, which approximates the vector field for the reduced dynamics as a neural network. The resulting low-dimensional data-driven model effectively captures short-time dynamics over the span of one correlation time, as well as long-time dynamics, particularly the self-similar, nested traveling wave structure of 2D EIT in the parameter range considered. 
%In the present study, we use data-driven methods to develop reduced-order models (ROM) of 2D EIT by evolving dynamics on the invariant manifold through neural ordinary differential equations which have much fewer degrees of freedom than the DNS. The dataset representing the dynamics of EIT has been mapped to the invariant manifold using a combination of a viscoelastic variant of proper orthogonal decomposition (VEPOD) and an encoding neural network, whereas the dynamics evolved on the manifold is mapped back to VEPOD space using a decoding neural network and then projected to the physical space. This ROM captures the dynamics of EIT $O(10^6)$ times faster than the DNS, accurately. We use short-time tracking and long-time statistics to show the quantitatively predictive accuracy of the model.    
\end{abstract}
%\begin{keywords}
%Authors should not enter keywords on the manuscript, as these must be chosen by the author during the online submission process and will then be added during the typesetting process (see http://journals.cambridge.org/data/\linebreak[3]relatedlink/jfm-\linebreak[3]keywords.pdf for the full list)
%\end{keywords}

\section{Introduction}
% This suggests that the dynamics of EIT can be represented by significantly lower DOF and hence a reduced-order model of the dynamics of EIT can be developed. 

Elastoinertial turbulence (EIT) is a chaotic state resulting from the interplay between inertia and elasticity, and is suspected to set a limit on the achievable drag reduction in turbulent flows using polymer additives \citep{Samanta2013,Shekar2019}. Direct numerical simulation (DNS) of EIT is computationally demanding due to the requirement of resolving small-scale dynamics, which are essential to sustain EIT \citep{Sid:2018gh}. The dynamics of EIT are fundamentally two-dimensional (2D) \citep{Sid:2018gh} and a 2D numerical simulation of EIT requires $O(10^6)$ degrees of freedom (DOF), making the investigation of its dynamics challenging. A reduced-order model (ROM) of EIT having fewer DOF would not only accelerate the investigation of the dynamics of EIT but may also open the door to developing control strategies to suppress EIT, which would allow turbulent drag reduction beyond the maximum drag reduction (MDR) limit \citep{Linot2023}.  It is also of fundamental interest for any complex flow phenomenon to know how many DOFs are actually required to describe its dynamics. Specifically, we have recently shown that the dynamics of two-dimensional EIT in channel flow are dominated by a self-similar family of well-structured traveling waves \citep{Kumar2024}, and a natural question is how many DOFs are required to capture this structure.

In this study, we use data-driven modeling techniques for the time evolution of 2D channel flow EIT. We begin by considering the full-state data $ \boldsymbol{q} $, which resides in the ambient space $ \mathbb{R}^{d_N} $ and evolves over time according to $d\boldsymbol{q}/dt = \boldsymbol{f}(\boldsymbol{q})$, where the mesh resolution and the number of state variables determine the size of $d_N$. The foundation of the data-driven reduced-order modeling approach applied here is that the long-time dynamics of a dissipative system collapse onto a relatively low-dimensional invariant manifold \citep{Hopf:1948bn,Foias:1988ux,Temam1989}. By mapping $ \boldsymbol{q} $ to invariant manifold coordinates $ \boldsymbol{h} \in \mathbb{R}^{d_h}  (d_h<d_N)$, we can describe the evolution of \( \boldsymbol{h} \) with a new equation $ d\boldsymbol{h}/dt = \boldsymbol{g}(\boldsymbol{h}) $ in these manifold coordinates. 

A classical approach to dimension reduction is Principal Components Analysis (PCA) (also known as Proper Orthogonal Decomposition or POD in fluid dynamics) \citep{Holmes_Lumley_Berkooz_Rowley_2012}. However, PCA projects data onto a flat manifold because it is an inherently linear technique, and thus may not adequately represent the generally non-flat invariant manifold where data from a complex nonlinear system lies. To capture nonlinear manifold structure, autoencoders are widely used \citep{Kramer1991,Milano2002,Linot2020}, which is the approach we pursue here. In high-dimensional systems, it can be beneficial to first apply PCA for linear dimension reduction, followed by using an autoencoder for further reduction \citep{Linot2023jfm,young2023scattering,pipe_jfm}. Once a low-dimensional representation of the full system state is identified, we can proceed with data-driven modeling of the dynamics to find $\boldsymbol{g}$.  We use the framework, known as `neural ODEs' (NODE) \citep{chen_node,Linot2022} that represents the vector field $\boldsymbol{g}$ as a neural network. It is important to note that the Markovian nature and continuous-time framework inherent to NODEs are well-aligned with the underlying physics of the nonlinear turbulent dynamics. This framework, which we denote DManD (Data-Driven Manifold Dynamics), has been successfully applied to a wide variety of nonlinear turbulent dynamics, including Kolmogorov flow, plane Couette flow, and pipe flow \citep{DeJesus2023,Linot2023jfm,pipe_jfm,Constante-Amores2024}. DManD models have proven capable of accurately tracking short-term dynamics for at least one Lyapunov time and capturing key statistics of long-term trajectories, such as Reynolds stresses and energy balance in wall-bounded flows. Additionally, these models have been used to aid the discovery of new exact coherent structures (ECSs), where the ECSs are direct solutions of the governing equations and they have well-defined structures that organize dynamics in chaotic flows (e.g., traveling waves, periodic orbits) \citep{Linot2023jfm,pipe_jfm}.

\begin{figure}
\centering
\includegraphics[width=\textwidth]{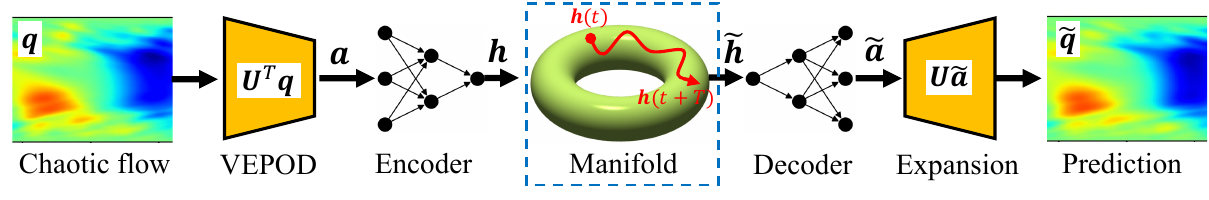}
\caption{Framework of Viscoelastic Data-driven Manifold Dynamics (VEDManD) used to develop a reduced-order model of elastoinertial turbulence.} 
\label{DManD_sketch.pdf}
\end{figure}

In this study, we develop a data-driven model of EIT in 2D channel flow. We introduce the Viscoelastic Data-driven Manifold Dynamics (VEDManD) framework, which enables the construction of low-dimensional models that faithfully capture the short-time tracking and long-time statistics in the present case using only 50 degrees of freedom (Fig. \ref{DManD_sketch.pdf}). The remainder of this article is organized as follows: Section 2 outlines the methodological framework, Section 3 discusses the results, and Section 4 concludes with final remarks.

\vspace{-5mm}

\section{Formulation and governing equations}
%\begin{comment}
%\end{comment}

\subsection{Direct numerical simulation of EIT}
% To develop a data-driven model of EIT, first, we need to generate the data describing the dynamics of EIT.
% We consider 2D EIT in the present study as their dynamics are fundamentally 2D \citep{Sid:2018gh}. 
The dimensionless momentum and incompressible mass conservation equations are 
\begin{equation}\label{colm}
\frac{\partial \boldsymbol{u}}{\partial t}+\boldsymbol{u}\cdot \nabla \boldsymbol{u} =-\nabla p+\frac{\beta}\Rey \nabla^2 \boldsymbol{u} + \frac{1-\beta}\Rey \nabla \cdot \boldsymbol{\tau}_p+f(t)\boldsymbol{e}_x,\qquad \nabla \cdot \boldsymbol{u}=0, 
\end{equation}
where the non-dimensional velocity and pressure fields are denoted by $\boldsymbol{u}$ and $p$, respectively. Lengths and velocities are made non-dimensional using the channel half-width ($H$) and the Newtonian laminar centerline velocity ($U_c$), respectively. The Reynolds number $\Rey=\rho U_c H/\eta$, where $\rho$ and $\eta$ are fluid density and total zero-shear rate viscosity. The solvent viscosity is $\eta_s$ and $\beta=\eta_s/\eta$. The polymer stress tensor $\boldsymbol{\tau}_p$ is modeled with the FENE-P constitutive equation: 
\begin{equation}\label{tau_theta}
\frac{\partial \boldsymbol{\alpha}}{\partial t}+\boldsymbol{u}\cdot \nabla \boldsymbol{\alpha} - \boldsymbol{\alpha}\cdot \nabla \boldsymbol{u}-(\boldsymbol{\alpha}\cdot \nabla \boldsymbol{u})^T=-\boldsymbol{\tau}_p+\frac{1}{\Rey \Sc} \nabla^2 \boldsymbol{\alpha},
\end{equation}
\begin{equation}\label{fenep}
\boldsymbol{\tau}_p= \frac{1}{\Wi} \left( \frac{\boldsymbol{\alpha}}{1-\mathrm{tr}(\boldsymbol{\alpha})/b}-\mathsfbi{I}\right),
\end{equation}
where $\boldsymbol{\alpha}$ is the conformation tensor, $b$ is the maximum extensibility of the polymer chains and $\mathsfbi{I}$ is the identity tensor. The Weissenberg number  $\Wi=\lambda U_c/H$, where $\lambda$ is the polymer relaxation time.  To ensure numerical stability, we include a small diffusion term in the evolution equation of the conformation tensor (Eq. \ref{tau_theta}), whose strength is controlled by the Schmidt number $Sc=\eta/\rho D$, where $D$ is the diffusion coefficient of the polymer molecules. 

We solve the governing equations in 2D channel flow using no-slip boundary conditions at the channel walls $y=\pm 1$. \textcolor{black}{The introduction of diffusion requires boundary conditions for $\boldsymbol{\alpha}$ to evolve Eq. \ref{tau_theta}. As has been done elsewhere (e.g. \cite{Sid:2018gh}), at the walls, we solve Eq. \ref{tau_theta} with no diffusion term (i.e., $1/Sc=0$) and use the solution as the boundary condition on $\boldsymbol{\alpha}$ for Eq. \ref{tau_theta} with finite $Sc$. } We impose periodic boundary conditions in the flow ($x$) direction. An external force in the streamwise direction ($f(t)\boldsymbol{e}_x$) drives the flow. This forcing is chosen at each time step to ensure that the volumetric flow rate remains at the Newtonian laminar value.
% which maintains the volumetric flow rate corresponding to the Newtonian laminar value.

Direct numerical simulations (DNS) are performed using \emph{Dedalus}  \citep{Burns2020}, an open-source tool based on the spectral method. The channel has a length of $5$ units and a height of $2$ units, and the computational domain is discretized using 256 Fourier and 1024 Chebyshev basis functions  \textcolor{black}{with 3/2 dealiasing factor} in the streamwise ($x$) and wall-normal ($y$) directions, respectively. We consider a dilute polymer solution ($\beta=0.97$) of long polymer chains ($b=6400$) at $\Rey=3000$,  $\Wi=35$. \textcolor{black}{For polymer additives, such as polyethylene oxide and polyacrylamide, this value of $b$ corresponds to molecular weights $\sim 300$ kDa and $\sim 500$ kDa, respectively.} The Schmidt number is $Sc=250$, consistent with previous studies \citep{Sid:2018gh,Kumar2024}.  

\vspace{-3mm}
\subsection{Linear dimension reduction: Viscoelastic Proper Orthogonal Decomposition}
Due to the very high dimension of the EIT data ($\approx 1.6 \times 10^6$), it is desirable to begin the dimension reduction process with a linear step to reduce to a smaller, (but still fairly large) dimension at which the subsequent nonlinear step will be more tractable (Fig. \ref{DManD_sketch.pdf}). To do so, we use a viscoelastic variant \citep{Wang2014} of POD \citep{Holmes_Lumley_Berkooz_Rowley_2012}, which we will denote `VEPOD'. The aim of (VE)POD is to find a function $\boldsymbol{\psi}(\boldsymbol{x})$ that maximizes the objective function 
\begin{equation}\label{objective_fn}
E\{ | \langle \boldsymbol{q}(\boldsymbol{x}), \boldsymbol{\psi}(\boldsymbol{x}) \rangle |^2\}
\end{equation}
given the constraint $\langle \boldsymbol{\psi}(\boldsymbol{x}), \boldsymbol{\psi}(\boldsymbol{x}) \rangle=1$, where $\boldsymbol{x}$ denotes the position and $\langle \cdot , \cdot\rangle$ represents an inner product (further discussed below). The expectation and modulus operations over the ensemble of data are given by $E\{\cdot\}$ and $|\cdot|$, respectively. Here, $\boldsymbol{q}(\boldsymbol{x})$ is a vector containing instantaneous state variables and the ensemble we average over is a long time series.

An appropriate inner product for VEPOD can be found by considering the total mechanical energy of the fluid, 
% The velocity field and polymer stress field in EIT have different magnitudes. Therefore,  it is essential to normalize both fields appropriately before organizing them in $\boldsymbol{q}(\boldsymbol{x})$. Appropriate normalization factors for the various components of the field can be obtained by considering the total mechanical energy of the EIT dynamics. In EIT, the total mechanical energy ($U_{tot}$) 
which consists of the kinetic energy contribution ($U_k$) from the velocity and the elastic energy contribution ($U_e$) from the stretching of the polymer chains. For the FENE-P model, precisely computing the mechanical energy is challenging because of the limitations introduced by Peterlin's approximation. However, an approximate mechanical energy can be given as:
\begin{equation}\label{mechanical_energy}
U_{tot}=U_{k}+U_{e}= \frac{1}{2} \int_{\Omega} \left\{\boldsymbol{u} \cdot \boldsymbol{u} + \frac{1-\beta}{Re Wi}\boldsymbol{\theta} : \boldsymbol{\theta}\right\}  d\boldsymbol{x}, 
\end{equation} 
where $\boldsymbol{\theta} \cdot \boldsymbol{\theta}=\boldsymbol{\alpha}/(1-\mathrm{tr}(\boldsymbol{\alpha})/b)$ \citep{Wang2014}. For the Oldroyd-B model ($b \to \infty$), Eq. \ref{mechanical_energy} represents the exact mechanical energy. Corresponding to this energy definition, an appropriate state variable $\boldsymbol{q}(\boldsymbol{x})$ is:
\begin{equation}\label{q_entry}
\boldsymbol{q}=\left[\boldsymbol{u}, \mathsfbi{T}\right],
\end{equation}
where $\mathsfbi{T}=\sqrt{\frac{1-\beta}{Re Wi}}\boldsymbol{\theta}$ is the weighted symmetric square root of the conformation tensor, which we call the stretch tensor. Consequently, the inner product used in Eq. \ref{objective_fn} can be defined as 
\begin{equation}\label{inner_product}
\langle\boldsymbol{q,\psi}\rangle= \int_{\Omega} \boldsymbol{q}(\boldsymbol{x}) \cdot \boldsymbol{\psi}(\boldsymbol{x}) d\boldsymbol{x}, 
\end{equation}
so that the inner product of $\boldsymbol{q}$ with itself yields $2U_{tot}$.
Maximization of the objective function (Eq. \ref{objective_fn}) gives a self-adjoint eigenvalue problem:
\begin{equation}\label{eigen_continous}
\int_{\Omega} E\{\boldsymbol{q}(\boldsymbol{x}) \boldsymbol{q}^*(\boldsymbol{y}) \}  \boldsymbol{\psi}(\boldsymbol{y})  d\boldsymbol{y}=\sigma  \boldsymbol{\psi}(\boldsymbol{x}),
\end{equation}
which leads to an infinite set of eigenmodes $\{\sigma_j, \boldsymbol{\psi}_j(\boldsymbol{x})\}$ arranged in decreasing value of the energy represented by the eigenvalue  $\sigma_j$. The eigenfunctions $\boldsymbol{\psi}_j(\boldsymbol{x})$ are orthonormal with the given inner product. The vector of VEPOD coefficients $\boldsymbol{a}$ is given by $a_j=\langle \boldsymbol{q}(\boldsymbol{x}), \boldsymbol{\psi}_j(\boldsymbol{x}) \rangle$. The state variables can be reconstructed using VEPOD modes as 
\begin{equation}\label{reconstruct_POD} \Tilde{\boldsymbol{q}}(\boldsymbol{x})=\sum_{j=1}^{d_a}a_j\boldsymbol{\psi}_j(\boldsymbol{x}),
\end{equation}
where $d_a$ represents the number of VEPOD modes used in the reconstruction. To estimate the VEPOD of a discrete dataset, we arrange the snapshots of the flow state as:
\begin{equation}\label{Q_snapshot} 
\mathsfbi{Q}=[\boldsymbol{q}_1, \boldsymbol{q}_2, ..., \boldsymbol{q}_{N_t}],
\end{equation}
where $\boldsymbol{q}_i=\boldsymbol{q}(t_i)$ and $N_t$ represents the total number of snapshots. The eigenmodes of VEPOD can be obtained by solving the following discretized eigenvalue problem:
\begin{equation}\label{eigenvalue_discretize} 
\mathsfbi{Q}\mathsfbi{Q}^*\mathsfbi{W}\boldsymbol{\psi}=\boldsymbol{\psi}\boldsymbol{\sigma},
\end{equation}
where $\boldsymbol{\sigma}$ is a diagonal matrix containing the eigenvalues, and $\mathsfbi{W}$ accounts for the numerical quadrature necessary for integration on a non-uniform grid.

\vspace{-3mm}
\subsection{Non-linear dimension reduction and NODE}
\begin{table}
\begin{center}
\def~{\hphantom{0}}
  \begin{tabular}{lccccc}
        Neural Network & Architecture & Activation  & Learning rate\\
      Encoder ($\boldsymbol{\mathcal{E}} $) & $4000:5000:1000:250:d_h$ & ReLU:ReLU:ReLU:Lin & [$4\times10^{-4}, 10^{-4}$] \\
      Decoder ($\boldsymbol{\mathcal{D}} $) & $d_h:250:1000:5000:4000$ & ReLU:ReLU:ReLU:Lin & [$4\times10^{-4}, 10^{-4}$]\\
      NODE ($\boldsymbol{g} $) & $d_h:500:500:d_h$ & Sig:Sig:Lin & [$10^{-4}, 10^{-5}$] \\
  \end{tabular}
  \caption{Details of different neural networks used in the VEDManD framework. `Architecture' represents the dimension of each layer and `Activation' refers to the types of activation functions used, where  `ReLU', `Sig', and `Lin' stand for Rectified Linear Unit, Sigmoid, and Linear activation functions, respectively. `Learning Rate' represents the learning rates used during training.}
  \label{table_nn}
  \end{center}
\end{table}

After projecting the data to the leading VEPOD modes, we perform a nonlinear dimension reduction with a `hybrid autoencoder' \citep{Linot2020}  to determine the mapping into the manifold coordinates $\boldsymbol{h}=\boldsymbol{\chi}(\boldsymbol{a})$, along with mapping back $\tilde{\boldsymbol{a}}=\check{\boldsymbol{\chi}}(\boldsymbol{h})$. This hybrid autoencoder uses two neural networks to learn the corrections from the leading VEPOD coefficients, as $\boldsymbol{h}=\boldsymbol{\chi}(\boldsymbol{a};\theta_\mathcal{E})=\mathsfbi{U}^T_{d_h} \boldsymbol{a}+\boldsymbol{\mathcal{E}}(\mathsfbi{U}^T_{d_a}\boldsymbol{a},\theta_\mathcal{E})$,
here, $\mathsfbi{U}_k \in \mathbb{R}^{d_N\times k}$ is a matrix whose columns are the first $k$ VEPOD eigenfunctions, and $\boldsymbol{\mathcal{E}}$ is an encoding neural network. 
The mapping back to the full space is given by
$\tilde{\boldsymbol{a}}=\check{\boldsymbol{\chi}}({\boldsymbol{h};\theta_\mathcal{E}})=\mathsfbi{U}_{d_a} ([\boldsymbol{h},0]^T+\boldsymbol{\mathcal{D}}(\boldsymbol{h};\theta_\mathcal{D}))$, where $\boldsymbol{\mathcal{D}}$ is a decoding neural network. The autoencoder minimizes the error 
\begin{equation}\label{loss_autoencoder} 
\mathcal{L}=||
 \boldsymbol{a}(t_i)-\check{\boldsymbol{\chi}}(\boldsymbol{\chi}(\boldsymbol{a}(t_i);\theta_\mathcal{E});\theta_\mathcal{D})||^2 +
 \kappa||
 \boldsymbol{\mathcal{E}}(\mathsfbi{U}^T_{d_a}\boldsymbol{a}(t_i);\theta_\mathcal{E})+\boldsymbol{\mathcal{D}}_{d_h}(\boldsymbol{h}(t_i);\theta_\mathcal{D})
 ||^2, 
\end{equation}
where the second term is a penalty to enhance the accurate representation of the leading $d_h$ VEPOD coefficients \citep{Linot2020} (in this study, $\kappa=1$). %\MDG{Somewhere we need to note that the losses are ensemble-averaged over training data.}
%So these functions can in principle reconstruct the state (i.e., $\boldsymbol{a}\approx \tilde{\boldsymbol{a}}$).} %\MDG{how? This needs a couple more details. Oh I see you have thsee below. Move the loss up to be part of this paragraph.}

% As mentioned earlier, the VEPOD truncation provides a linear dimension reduction and it cannot map dynamics on a curved invariant manifold. Therefore, we use the VEPOD coefficients ($\boldsymbol{a}$) for further dimension reduction using nonlinear methods. We use a neural network called hybrid autoencoder to discover the minimal dimension representation of the flow state and map back to the VEPOD representation of the state. Hence, the autoencoder has two parts: Encoder ($\boldsymbol{\mathbb{E}}$) and Decoder ($\boldsymbol{\mathbb{D}}$). The encoder maps the VEPOD representation of the state to the invariant manifold representation as
% \begin{equation}\label{encoder_nn} 
% \boldsymbol{h}=\boldsymbol{\mathbb{E}(a)},
% \end{equation}
% where $\boldsymbol{h}$ denotes the invariant manifold representation of the full state that has a significantly lower dimension ($d_h=O(10)$). 
% The decoder $\boldsymbol{\mathbb{D}}$ maps back the flow state from manifold representation to VEPOD representation:
% \begin{equation}\label{decoder_nn} 
% \boldsymbol{\Tilde{a}}=\boldsymbol{\mathbb{D}(h)},
% \end{equation}
%where $\boldsymbol{\Tilde{a}}$ represents the autoencoder reconstruction of the state. 
Once the low-dimensional representation of the full state is discovered, we use a `stabilized' NODE to learn the dynamics in manifold coordinates: i.e.~ 
$ d\boldsymbol{h}/dt=\boldsymbol{g} (\boldsymbol{h})-\mathsfbi{A}\boldsymbol{h}$,
where $\boldsymbol{g}$ is a neural network, and  $\mathsfbi{A}=\gamma \mathsfbi{I} S(\boldsymbol{h})$ is a diagonal matrix, where $S(\boldsymbol{h})$ is the standard deviation of $\boldsymbol{h}$ and $\gamma$ is a tunable parameter. This term stabilizes the system by preventing the dynamics from drifting away from the attractor \citep{Linot2023jcp}.  The NODE is trained to minimize the difference between the true state $\boldsymbol{h}(t_i+\Delta t_s)$ and the predicted state $\tilde{\boldsymbol{h}}(t_i+\Delta t_s)$:
\begin{equation}\label{loss_NODE} 
\mathcal{J}=||\boldsymbol{\Tilde{h}}(t_i+\Delta t_s)-\boldsymbol{h}(t_i+\Delta t_s)||^2,
\end{equation}
where $\boldsymbol{\Tilde{h}}(t_i+\Delta t_s)=\boldsymbol{h}(t_i)+\int_{t_i}^{t_i+\Delta t_s} \left(\boldsymbol{g} (\boldsymbol{h}) - \mathsfbi{A}\boldsymbol{h}\right)dt$ is a time forward prediction of the NODE over time interval $\Delta t_s$. Details of the different neural networks are summarised in Table \ref{table_nn}.
% Thus, our Viscoelastic Data-driven Manifold Dynamics (VEDManD) model of EIT contains three different neural networks and their details have been given in Table \ref{table_nn}.

The ROM of EIT in the present study has been developed using a statistically stationary dataset consisting of $1000$ time units simulated using time step $\Delta t=0.001$ and sampled at the interval of $\Delta t_s=0.025$ time unit, which leads to a total of $40000$ snapshots. The correlation time of the dynamics of EIT is small ($<1$ time unit), so this dataset is sufficient to develop our ROM. Here, $80 \%$ of data have been used for training and $20 \%$ data for testing. The losses used to train the neural networks (Eqs. \ref{loss_autoencoder} and \ref{loss_NODE}) are the ensemble-averaged over the training data. \textcolor{black}{The DNS was performed in parallel mode on 24 processors on an HPC cluster and it took $\approx 2$ weeks to compute for $1000$ time units. The VEDManD framework consists of 3 steps, and a single processor was used to develop our ROM. The first step, the projection of the dataset on $4000$ VEPOD modes, takes $\approx 4$ hours. The second step, the training of the autoencoder, takes $\approx2$ days.  The third step, the training of the NODE, takes $2-3$ hours. Thus, the development of ROM takes $\approx2.5$ days on a single processor. Once, the model is developed, it evolves dynamics for $1000$ time units in seconds on a single processor.} 
%The autoencoder (500 epochs) and NODE (400000 epochs) have been trained until their respective training losses plateau.

\vspace{-5mm}
\section{Results and discussion}
\vspace{-1mm}
\subsection{Dimension reduction of EIT} 
\begin{figure}
\centering
\includegraphics[width=\textwidth]{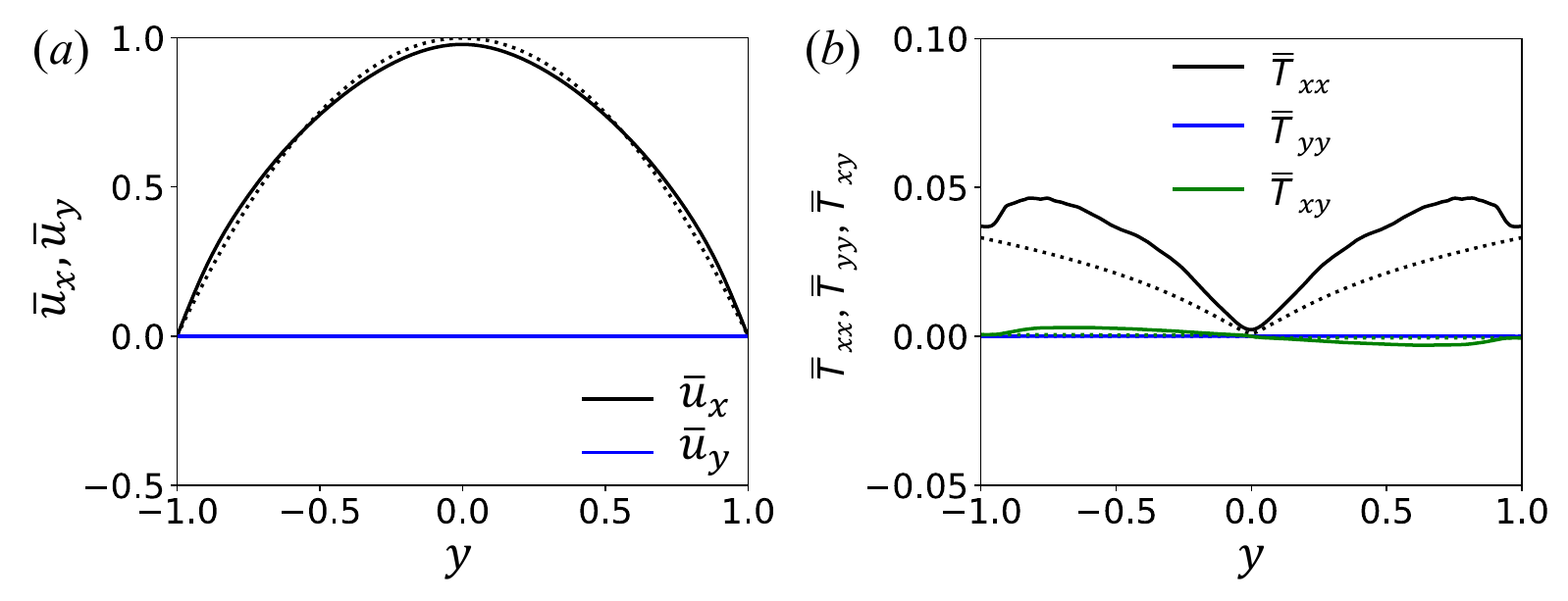}
\caption{Mean profiles (solid lines) of the components of ($a$) velocity and ($b$) stretch tensor in EIT at $\Rey =3000$ and $\Wi=35$. The dotted lines show the laminar profiles at the same parameter. \textcolor{black}{The temporal mean profiles of velocity in EIT are close to the laminar profiles as velocity fluctuations in EIT are weak \citep{Sid:2018gh}.}}
\label{velocity_stress_mean_laminar_dns_wi35.pdf}
\end{figure}

\vspace{-1mm}
For statistically stationary dynamics, the instantaneous state variables can be written as $\boldsymbol{u}=\overline{\boldsymbol{u}}+\boldsymbol{u}^{\prime}$ and $\mathsfbi{T}=\overline{\mathsfbi{T}}+\mathsfbi{T}^{\prime}$, where $\overline{(\cdot)}$ and $(\cdot)^{\prime}$ are the temporal mean and perturbation, respectively. Mean profiles and laminar profiles are shown in Fig. \ref{velocity_stress_mean_laminar_dns_wi35.pdf}$a$ and Fig. \ref{velocity_stress_mean_laminar_dns_wi35.pdf}$b$, respectively. Here, we develop a ROM to capture the perturbations of the state variables: $\boldsymbol{q}=[\boldsymbol{u}',\mathsfbi{T}']$.

\begin{figure}
\centering
\includegraphics[width=\textwidth]{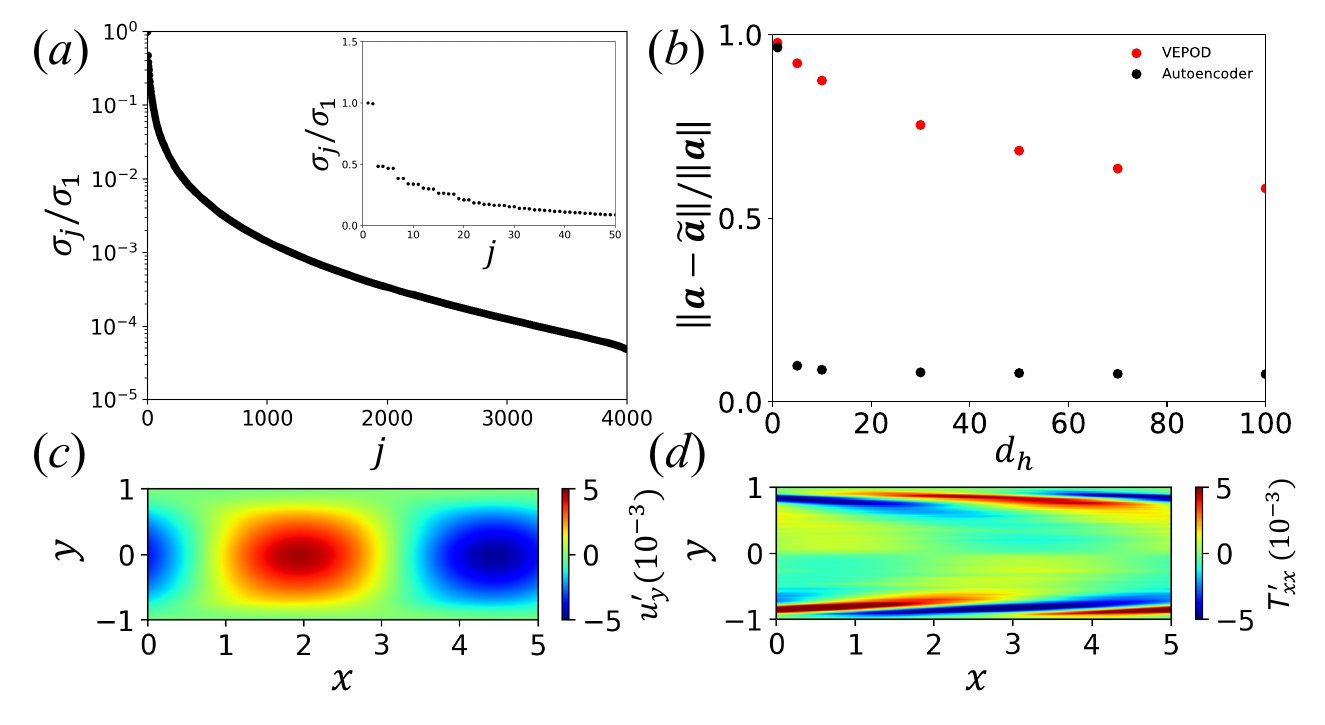}
\caption{($a$) VEPOD eigenvalue spectrum. %Reconstruction of the instantaneous profiles of components of ($b$) velocity and ($c$) stretch tensor at $x=2.5$ from DNS (solid lines) and VEPOD with 4000 modes (dashed lines). 
($b$) Normalized reconstruction error on the test dataset for various latent dimensions of VEPOD and autoencoder. \textcolor{black}{Leading VEPOD mode structure for ($c$) $u'_y$ and ($d$) $T_{xx}'$.}}
\label{dimension_reduction_POD_AE.pdf}
\end{figure}

The VEPOD eigenvalue spectrum of the perturbations in state variables, Fig. \ref{dimension_reduction_POD_AE.pdf}$a$, shows that the energy content of higher VEPOD modes decreases approximately exponentially for index $j\gtrsim 1500$. We find that reconstruction of the flow state using $d_a=4000$ VEPOD modes captures $\approx 99.8 \%$ of the total mechanical energy and yields an accurate representation. Therefore, to develop our model we retain $4000$ VEPOD modes. \textcolor{black}{We also visualize the leading VEPOD mode structure for $u_y'$ and $T_{xx}^{\prime}$ (Figs. \ref{dimension_reduction_POD_AE.pdf}(c,d)) and find that they resemble the most dominant traveling wave underlying EIT \citep{Kumar2024}.}

\begin{figure}
\centering
\includegraphics[width=\textwidth]{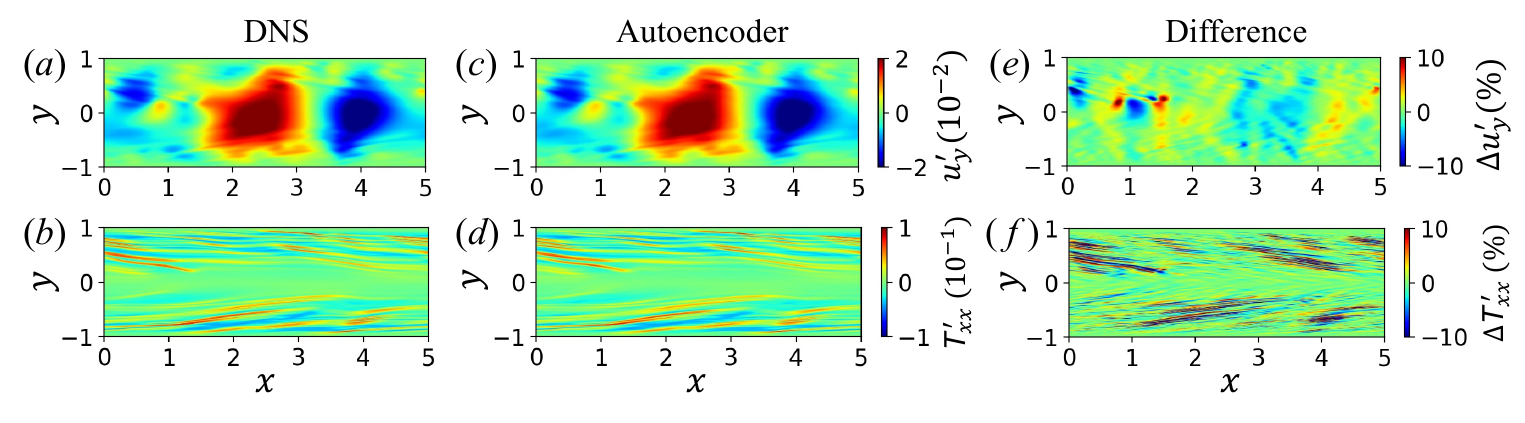}
\caption{Instantaneous ($a,c$) $u'_y$ and ($b,d$) $T_{xx}'$ obtained from ($a,b$) DNS and ($c,d$) reconstruction with an autoencoder with $d_h=50$. ($e,f$) Difference between DNS and autoencoder reconstruction normalized with the maximum values of respective fields. %Note that the scales of $\Delta u'_y$ and $\Delta T_{xx}'$ are one order of magnitude lower than that of $u'_y$ and $T_{xx}'$. %\MDG{Times font}
% of different state variables of the flow state.
}
\label{dns_pod_autoencoder_flow_state.pdf}
\end{figure}

Before training the neural networks,  we first center and scale the dataset by subtracting the mean and then normalizing it with the maximum standard deviation. To find a low-dimensional representation of the system, we train multiple autoencoders with varying numbers of latent dimensions and visualize the reconstruction error of VEPOD coefficients on the test dataset (Fig. \ref{dimension_reduction_POD_AE.pdf}$b$). The reconstruction error initially decreases rapidly at low $d_h$, then shows only a marginal improvement in the reconstruction for latent dimension $d_h>30$. The VEDManD model with $d_h=30$ struggles to quantitatively capture the dynamics, so we do not report results for it and we use the latent dimension of the autoencoder $d_h=50$ in the present study. To illustrate the accuracy of the reconstruction, we compare results for $u_y'$ and $T_{xx}^{\prime}$ fields using the autoencoder at $d_h=50$ (Fig. \ref{dns_pod_autoencoder_flow_state.pdf}($c,d$)) with the corresponding DNS results (Fig. \ref{dns_pod_autoencoder_flow_state.pdf} ($a,b$)). The autoencoder faithfully reconstructs the flow state obtained from the DNS. 
Visual inspection of the autoencoder reconstruction reveals a very close resemblance, while a quantitative comparison between DNS and autoencoder results demonstrates differences at fine scales in the reconstruction ($\sim 10 \%$ for the worst-case scenario and are confined only in a tiny region), resulting from the cumulative effects of VEPOD truncation and autoencoder reconstruction (Fig. \ref{dns_pod_autoencoder_flow_state.pdf} ($e,f$)). Further analysis shows that the error is dominated by VEPOD truncation to 4000 modes -- the autoencoder error is very small. The other components exhibit similar degrees of accuracy.

% Once the autoencoder maps the system on a low-dimensional representation,
% % invariant manifold,
% we train the NODE to predict dynamics.
% on the manifold. The training of NODE is significantly faster than the autoencoder as it has a shallower and thinner architecture due to fewer input dimensions compared to the autoencoder. 
%
% For any arbitrary initial condition in the physical space, the corresponding initial condition for the NODE in the latent space can be obtained by using VEPOD truncation and encoder. Then, a time series of the dynamics on the invariant manifold can be generated using the NODE, which can be mapped back to the VEPOD coefficients using the decoder and then can be projected to the physical space using Eq. \ref{reconstruct_POD}. The NODE generates a time series of the dynamics of EIT significantly faster than the DNS for two specific reasons: (i) a smaller degree of freedom and (ii) a larger time step for temporal evolution. The neural ODE has just $50$ degrees of freedom, which is substantially lower than $1.6\times 10^6$ required to perform DNS. The time step in DNS depends on the mesh resolution and we have used a time step $\Delta t=0.001$ to successfully evolve the dynamics in DNS. However, the NODE can be evolved with a time step equivalent to the sampling time of the dataset used during the training, which is $\Delta t_s =0.025$ time unit in the present study. Thus, the VEDManD model of EIT allows us to generate the time series of dynamics of EIT $O(10^6)$ times faster than the DNS.   

% \subsection{Short-time tracking}
\vspace{-4mm}
\subsection{Low-dimensional dynamic model}
\begin{comment}
\begin{figure}
\centering
\begin{subfigure}[b]{0.32\textwidth} 
\includegraphics[width=\textwidth]{time_corralation_pod_dmand_tinit1840_tau200_wi35_da4000.pdf}
\caption{}
\label{time_corralation_pod_dmand}
\end{subfigure}
\begin{subfigure}[b]{0.32\textwidth} 
\includegraphics[width=\textwidth]{short_time_tracking_a_multiple_ICs_0_4_scaled_da4000.pdf}
\caption{}
\label{short_time_tracking_a_ex631_0_5_scaled}
\end{subfigure}
\begin{subfigure}[b]{0.32\textwidth} 
\includegraphics[width=\textwidth]{short_time_tracks_avg_a_N100_scaled_da4000.pdf}
\caption{}
\label{short_time_tracks_avg_a_N100}
\end{subfigure}
        \caption{(a) Temporal autocorrelation of the VEPOD coefficients. (b) Norm of the VEPOD coefficients for four arbitrary initial conditions using DNS and VEDManD with $d_h =50$ up to $t/t_c=4$. (c) Ensemble-averaged relative error between the VEPOD coefficients obtained using DNS and VEDManD.}
\end{figure}
\end{comment}

\begin{figure}
\centering
\includegraphics[width=\textwidth]{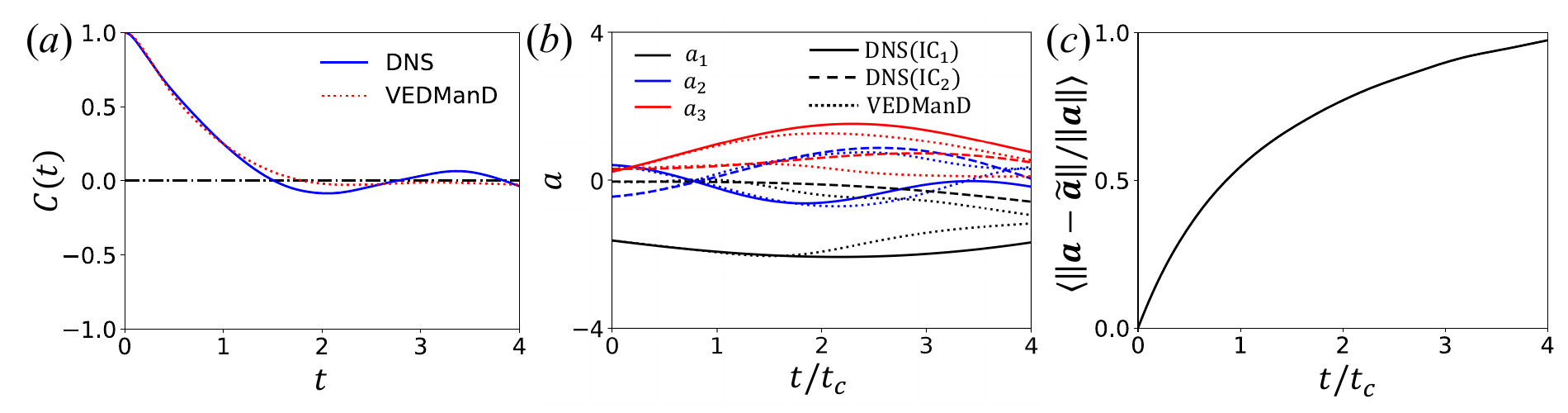}
\caption{($a$) Temporal autocorrelation of the VEPOD coefficients. ($b$) \textcolor{black}{First several VEPOD coefficients up to $t/t_c=4$ obtained using DNS and VEDManD with $d_h =50$ for two arbitrary initial conditions (IC$_1$ and IC$_2$).} ($c$) Ensemble-averaged relative error between the VEPOD coefficients obtained using DNS and VEDManD.}
\label{short_time_tracking.pdf}
\end{figure}

\begin{figure}
\centering
\includegraphics[width=\textwidth]{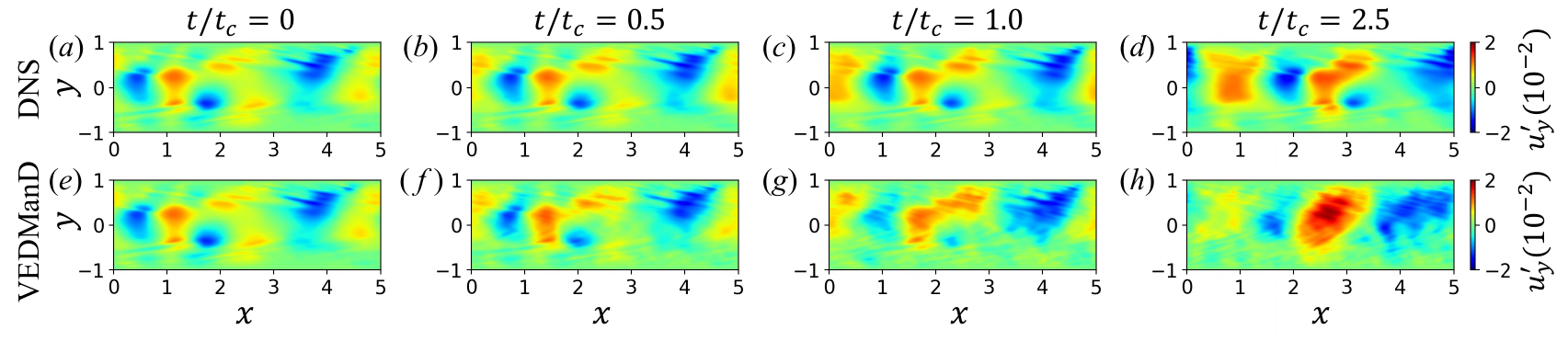}
\caption{Time series of $u_y^{\prime}$ obtained using ($a-d$) DNS and ($e-h$) VEDManD \textcolor{black}{for IC$_1$ in Fig. \ref{short_time_tracking.pdf}$b$}.}%Time series of ($a-h$) $u_y^{\prime}$ and ($i-p$) $T_{xx}^{\prime}$ obtained using ($a-d, i-l$) DNS and ($e-h, m-p$) VEDManD for IC$_1$ in Fig. \ref{short_time_tracking.pdf}$b$.
\label{uy_stress_short_time_prediction.pdf}
\end{figure}

Once the autoencoder maps the system on a low-dimensional representation, we train the NODE to predict dynamics. We start the presentation of the NODE results with short-time tracking. It is expected that the VEDManD predictions should closely follow DNS trajectories over a short time period before the dynamics diverge due to the chaotic nature of the system. To estimate the correlation time of the dynamics, we defined a temporal autocorrelation function ($C(t)$) for the VEPOD coefficients as: 
% \begin{equation}\label{temporal_corr} 
$C(t)=\frac{\langle \boldsymbol{R}(\tau) \cdot \boldsymbol{R}(\tau+t)\rangle}{\langle \boldsymbol{R}(\tau) \cdot \boldsymbol{R}(\tau)\rangle},$
%\end{equation}
where $\langle \cdot \rangle$ represents ensemble average. Here, the parameter $\boldsymbol{R}$ stands for $\boldsymbol{a}$ and $\Tilde{\boldsymbol{a}}$ for DNS and VEDManD, respectively. The autocorrelation functions from DNS and VEDManD are plotted in Fig. \ref{short_time_tracking.pdf}$a$. The correlation time of the dynamics is defined as $t_c=\int_0^{t_z}C(t) dt$, where $t_z$ represents the first zero of the autocorrelation function. The computed values using DNS and VEDManD are $t_c=0.67$ and $t_c=0.68$ respectively, very close to each other. %which demonstrate that the VEDManD model is capable of capturing the time-series of the dynamics over one correlation time \MDG{No that's not what it demonstrates -- you haven't shown any actual trajectory comparisons yet. It shows that the DManD accurately captures the correlation function}. 
This time scale indicates when trajectories starting from different initial conditions or with slightly different evolution equations should start to diverge from one another. In Fig.~\ref{short_time_tracking.pdf}$b$, we show \textcolor{black}{the time evolution of the first several VEPOD coefficients} for different initial conditions on the attractor using the VEDManD model and DNS. For $t<t_c$, the model accurately predicts the EIT trajectory, deviating at longer times. In  Fig. \ref{uy_stress_short_time_prediction.pdf}, we compare snapshots of the state obtained from the model against DNS with the same initial condition and see that for $t<t_c$ the results are very close. To quantify the deviation of the VEDManD prediction of trajectories from the DNS results, we show in Fig. \ref{short_time_tracking.pdf}$c$ the ensemble average of the relative error of the VEPOD coefficients from VEDManD and DNS calculated over many initial conditions. The difference between the model prediction and the DNS result is small at short times, increasing on the time scale $t_c$ before saturating. In summary, the VEDManD model with $d_h=50$ is capable of accurately capturing the dynamics of 2D EIT over time scales comparable to the flow correlation time.

% \subsection{Long-time statistics}
\begin{comment}
\begin{figure}
\centering
\begin{subfigure}[b]{0.48\textwidth} 
\includegraphics[width=\textwidth]{mode_energy_vs_frequency_uy_Re3000_Wi35_nfft2000_fm12k_dns_wi35.pdf}
\caption{}
\label{spod_dns}
\end{subfigure}
\begin{subfigure}[b]{0.48\textwidth} 
\includegraphics[width=\textwidth]{mode_energy_vs_frequency_uy_Re3000_Wi35_nfft2000_fm12k_dt025_t12k_24k_wi35_da4000.pdf}
\caption{}
\label{spod_dmand_a10}
\end{subfigure}
\caption{SPOD eigenvalue spectra of $u_y^{\prime}$ obtained using (a) DNS and (b) VEDManD. Red symbols indicate peaks in the leading mode of the spectra.}\label{fig:SPODspectra}
\end{figure}
\end{comment}

\begin{figure}
\centering
\includegraphics[width=\textwidth]{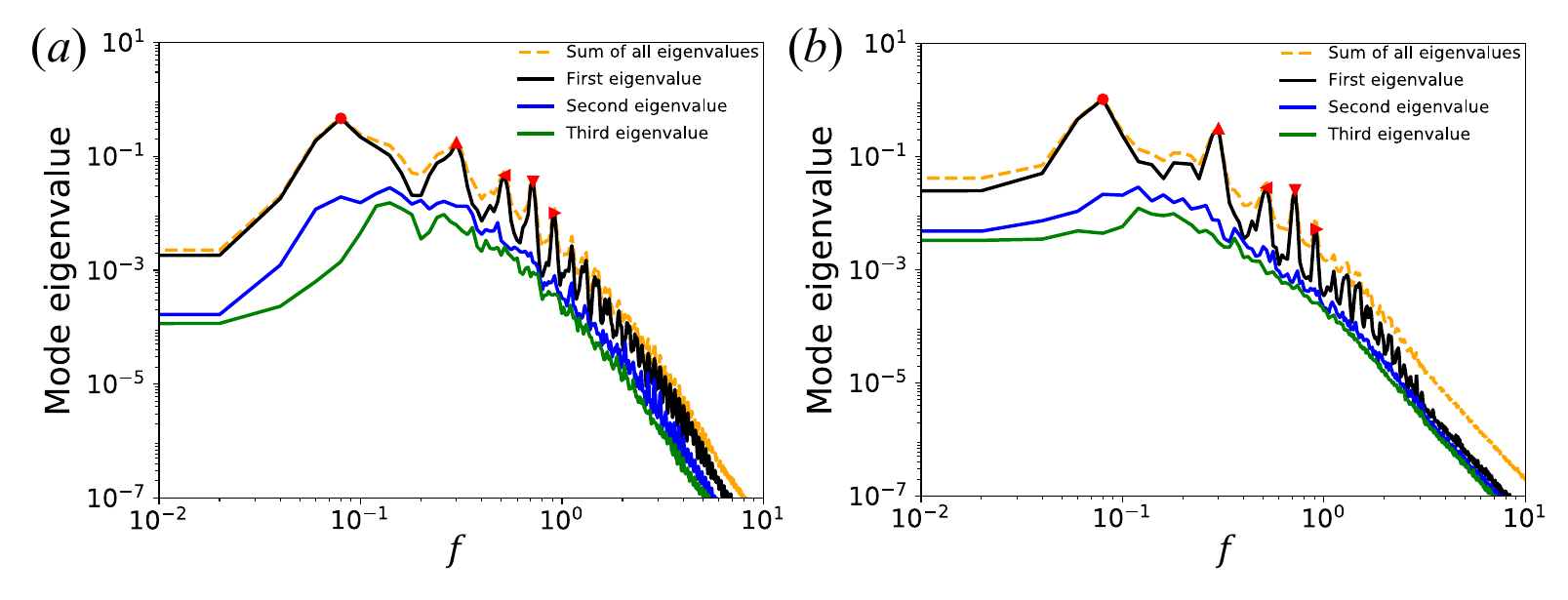}
\caption{SPOD eigenvalue spectra of $u_y^{\prime}$ obtained using ($a$) DNS and ($b$) VEDManD. Red symbols indicate peaks in the leading mode of the spectra.}
\label{spod_dns_vedmand_wi35.pdf}
\end{figure}

\begin{figure}
\centering
\includegraphics[width=\textwidth]{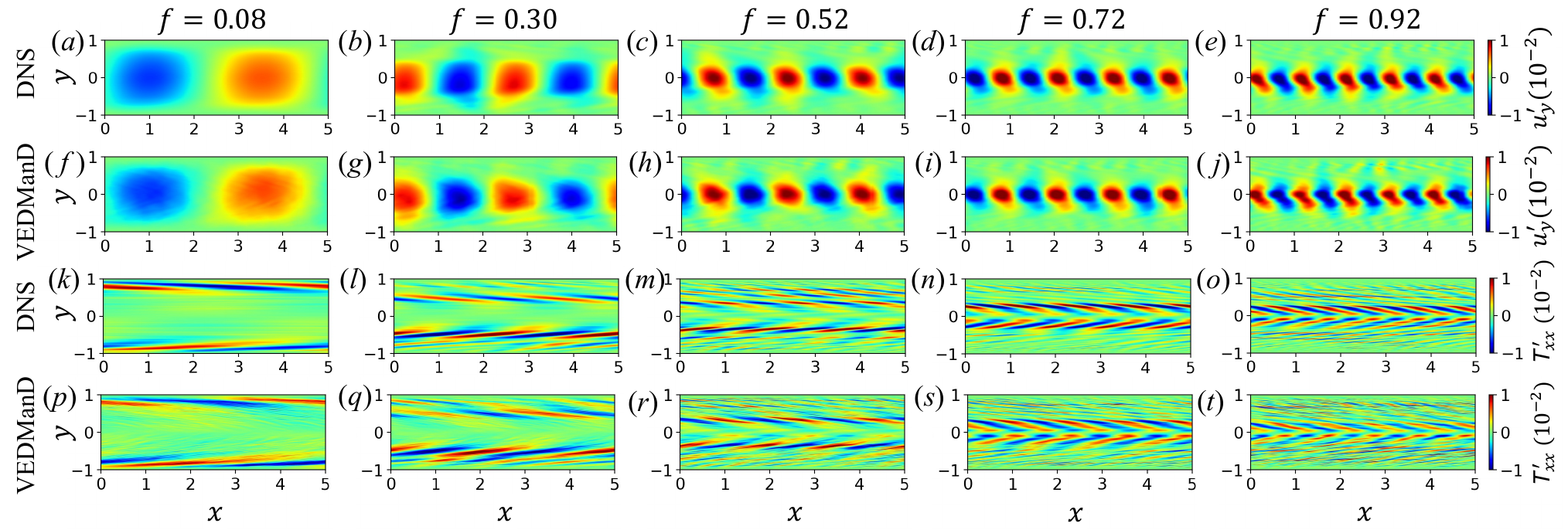}
\caption{SPOD mode structures of ($a-j$) $u_y^{\prime}$ and  ($k-t$) $T_{xx}^{\prime}$ from ($a-e,k-o$) DNS and ($f-j,p-t$) VEDManD at the frequencies indicated with red symbols in Fig.~\ref{spod_dns_vedmand_wi35.pdf}.}
\label{uy_spod_mode_structures_dns_dmand}
\end{figure}

We now turn to the long-time statistics of the time-evolution, focusing on the Spectral Proper Orthogonal Decomposition (SPOD) \citep{Towne2018}, of the flow and stretch fields. \textcolor{black}{In short, SPOD is a temporal variant of POD that seeks the frequency-by-frequency POD of a Fourier-transformed time-dependent flow, yielding for every frequency $f$ a spectrum of energies and modes.} For a detailed description of SPOD in the context of EIT, see \citet{Kumar2024}, where we used SPOD to show that the dynamics of 2D EIT in the parameter regime examined are dominated by a family of self-similar, nested traveling waves. Here, we report the SPOD spectrum of $u_y^{\prime}$, as this component gives the cleanest spectrum. The SPOD energy spectra of $u_y^{\prime}$ obtained from DNS and VEDManD are shown in Fig. \ref{spod_dns_vedmand_wi35.pdf}$a$ and Fig. \ref{spod_dns_vedmand_wi35.pdf}$b$, respectively. The leading modes of both spectra contain most of the energy and exhibit peaks at the same frequencies, at least to the two significant digits (red symbols) which is the frequency resolution in the present study. We visualize the SPOD mode structures corresponding to the peaks in the SPOD spectra in Fig. \ref{uy_spod_mode_structures_dns_dmand}. The structures obtained from the VEDManD model closely resemble the structures of traveling modes obtained from the DNS. \textcolor{black}{(The mode structures are traveling waves, therefore we align their spatial phases to make comparison easier.) The precise prediction of wavenumbers and frequencies of different mode structures by the VEDManD model gives an accurate prediction of their wave speeds.} Thus, we have found a highly accurate representation of EIT with $O(10)$ dimensions commencing from an initial state dimension of $O(10^6)$. This dimension might not be the true dimension of the invariant manifold where the dynamics live (this quantity remains difficult to precisely estimate from data for complex chaotic systems), but it is capable of providing models that faithfully capture the short-time dynamics and long-time statistics of elastoinertial turbulence.% $O(10^6)$ times faster than DNS.} %\MDG{align  them to make the comparison easier (and say that you have done so)} and they do not represent any shortcomings of the model. %The SPOD analyses of other components of state variables lead to the same conclusion.

%\MDG{DONE TO HERE 8/21}
\vspace{-7mm}
\section{Conclusion}
Numerical simulations of elastoinertial turbulence (EIT) are computationally expensive due to the high number of degrees of freedom ($O(10^6)$) required to resolve all spatial and temporal scales. In the present study, we use data-driven methods to develop a reduced-order model (ROM) that has far fewer degrees of freedom ($O(10)$) yet nevertheless captures these scales and the dynamics of EIT. To find the low dimensional representation of the full state, we first introduce a variant of proper orthogonal decomposition (VEPOD) which reduces the dimension of the EIT dataset from $O(10^6)$ to $O(10^3)$, and then use autoencoders to perform nonlinear dimension reduction from $O(10^3)$ to $O(10)$. To evolve dynamics on this low-dimensional representation, we use stabilized neural ordinary differential equations. A ROM with a dimension of 50 accurately predicts both short-time dynamics and long-time statistics. This model successfully captures the trajectory of the dynamics over the span of one correlation time. To analyze the long-time statistics of the dynamics, we use spectral proper orthogonal decomposition (SPOD) and show that the ROM accurately captures the complex nonlinear structures and the frequencies of the self-similar traveling waves that underlie the chaotic dynamics observed in the DNS of 2D EIT.

By accurately modeling EIT with significantly fewer degrees of freedom than DNS requires, manifold dynamics models like those presented here make it possible to perform computationally efficient analyses, such as calculating Floquet multipliers and local Lyapunov exponents. \textcolor{black}{Such models can accelerate the discovery of new exact coherent structures (e.g., periodic orbits, relative periodic orbits) underlying EIT \citep{Linot2023jfm}. These models could also facilitate the design of control strategies (such as patterning surfaces) using a reduced number of degrees of freedom which would allow turbulent drag reduction using polymer additives beyond the maximum drag reduction limit.}  

%\vspace{-3mm}

\noindent \textbf{Acknowledgments.} This research was supported under grant ONR N00014-18-1-2865 (Vannevar Bush Faculty Fellowship).

\noindent \textbf{Declaration of Interests.} The authors report no conflict of interest. 
% The Ref. \citep{Kumar2024} is to appear in the Journal of Fluid Mechanics. 

\vspace{-5mm}
\bibliographystyle{jfm}
\bibliography{main-nested,turbulence-MDG-2402}

\end{document}